\documentclass[12pt,english,pra, reprint]{revtex4-1}
\usepackage[T1]{fontenc}
\usepackage[latin9]{inputenc}
\usepackage{refstyle}
\usepackage{amsbsy}
\usepackage{amstext}
\usepackage{amssymb}
\usepackage{graphicx}

\makeatletter

\AtBeginDocument{\providecommand\secref[1]{\ref{sec:#1}}}
\AtBeginDocument{\providecommand\tabref[1]{\ref{tab:#1}}}
\AtBeginDocument{\providecommand\figref[1]{\ref{fig:#1}}}
\providecommand{\tabularnewline}{\\}
\RS@ifundefined{subsecref}
  {\newref{subsec}{name = \RSsectxt}}
  {}
\RS@ifundefined{thmref}
  {\def\RSthmtxt{theorem~}\newref{thm}{name = \RSthmtxt}}
  {}
\RS@ifundefined{lemref}
  {\def\RSlemtxt{lemma~}\newref{lem}{name = \RSlemtxt}}
  {}

\usepackage{braket}
\usepackage[version=3]{mhchem}
\usepackage{url}
\usepackage{notes2bib}

\makeatother

\usepackage{babel}
\begin{document}
\newcommand*\citeref[1]{\cite{#1}}
\newcommand*\citerefs[1]{\cite{#1}} 

\newcommand*\Erkale{{\sc Erkale}}
\newcommand*\HelFEM{{\sc HelFEM}}
\newcommand*\eg{\emph{e.g.}}
\newcommand*\ie{\emph{i.e.}}
\title{On the accurate reproduction of strongly repulsive interatomic potentials}
\author{Susi Lehtola}
\affiliation{Department of Chemistry, University of Helsinki, P.O. Box 55 (A. I. Virtasen
aukio 1), FI-00014 University of Helsinki, Finland}
\email{susi.lehtola@alumni.helsinki.fi}

\selectlanguage{english}%
\begin{abstract}
Knowledge of the repulsive behavior of potential energy curves $V(R)$
at $R\to0$ is necessary for understanding and modeling irradiation
processes of practical interest. $V(R)$ is in principle straightforward
to obtain from electronic structure calculations; however, commonly-used
numerical approaches for electronic structure calculations break down
in the strongly repulsive region due to the closeness of the nuclei.
In the present work, we show by comparison to fully numerical reference
values that a recently developed procedure {[}S. Lehtola, J. Chem.
Phys. 151, 241102 (2019){]} can be employed to enable accurate linear
combination of atomic orbitals calculations of $V(R)$ even at small
$R$ by a study of the seven nuclear reactions \ce{He2 <=> Be}, \ce{HeNe <=> Mg},
\ce{Ne2 <=> Ca}, \ce{HeAr <=> Ca}, \ce{MgAr <=> Zn}, \ce{Ar2 <=> Kr},
and \ce{NeCa <=> Zn}.
\end{abstract}
\maketitle
\global\long\def\ERI#1#2{(#1|#2)}%
\global\long\def\bra#1{\Bra{#1}}%
\global\long\def\ket#1{\Ket{#1}}%
\global\long\def\braket#1{\Braket{#1}}%

\section{Introduction\label{sec:Introduction}}

The interaction of high-energy particles with matter is typically
modeled using pairwise potentials (see \eg{} chapter 6 of \citeref{Sigmund2014}),
as the dominant interactions are determined by the highly repulsive
nuclear Coulomb barriers that are pairwise terms; see \eg{} \citeref{Karolewski2007}
for a recent numerical demonstration for low-energy projectiles incident
on copper surfaces. Most practical simulations employ the universal
potential by Ziegler, Biersack and Littmark \citep{Ziegler1985} (ZBL)
which is based on Thomas--Fermi calculations of the repulsive barrier.
However, Thomas--Fermi theory has significant shortcomings; for instance,
it is well known not to bind any molecules, and a method lacking these
shortcomings like Hartree--Fock (HF) or density-functional theory
\citep{Hohenberg1964,Kohn1965} (DFT) would certainly be more attractive.

\emph{Ab initio }calculations of the diatomic potential energy curve
(PEC), denoted here as $V_{AB}(R)$, are, however, challenging at
small internuclear distances $R$ due to the closeness of the two
nuclei. In contrast to chemistry at ambient conditions, even the innermost
core electrons may be significantly affected by the interaction between
the two atoms: for instance, in the \ce{Ar2 <=> Kr} nuclear reaction
obtained as $R\to0$, the \emph{two} $\text{[Ne]}3s^{2}3p^{6}$ electronic
configurations of the argon atoms deform into the \emph{single} $\text{[Ne]}3s^{2}3p^{6}4s^{2}3d^{10}4p^{6}$
configuration of the krypton atom. An extremely flexible numerical
approach must be used in order to describe such changes accurately,
obviously disallowing the use of pseudopotential and frozen-core approaches.
Although some efforts for the \emph{ab initio} description of the
Coulomb barrier have been made in the literature (see \eg{} \citerefs{Karolewski2007}, \cite{Sabelli1978, Sabelli1979, Keinonen1991, Keinonen1994, Nordlund1997, Pruneda2004, Kuzmin2006, Karolewski2006, Kuzmin2007, Juslin2008, Karolewski2012, Zinoviev2017}
and references therein), the problem of facile computation of $V_{AB}(R)$
for $R\to0$ remains still unsolved in the general case.

All-electron calculations are typically undertaken within the linear
combination of atomic orbitals (LCAO) approach. However, also the
LCAO approach fails in this case, because the basis functions on the
atoms $A$ and $B$ quickly become linearly dependent when $R\to0$.
Moreover, large atomic basis sets should be used in order to allow
the necessary flexibility for the core orbitals to deform in presence
of the other nucleus and its electrons. But, the more functions are
included in the calculations, the more linear dependencies are generated
when the nuclei start coinciding, and the calculations become numerically
unstable as the basis set becomes ill-behaved.

As always, fully numerical electronic structure calculations are one
option, see \citeref{Lehtola2019c} for a recent review. Here, the
numerical basis set can always be chosen in such a way that linear
dependencies do not arise even at small $R$. However, fully numerical
approaches carry a much higher computational cost than that of LCAO
calculations using \eg{} Gaussian basis sets, and may also be harder
to set up; see the discussion in \citerefs{Lehtola2019c, Lehtola2019b}.
Moreover, fully numerical electronic structure programs are less-developed
than Gaussian-basis ones, because the huge number of basis functions
in a fully numerical approach may \eg{} make sophisticated convergence
algorithms intractable \citep{Lehtola2019c}, making it more difficult
to carry out the wanted electronic structure calculations.

Despite the numerical problems encountered in standard LCAO approaches,
it should be perfectly well possible to describe diatomic molecules
using atomic basis sets even at small internuclear distances, because
at small $R$ the molecule looks like the compound atom that is especially
easy to describe with atomic basis sets. This means that the problems
in LCAO calculations should be circumventable by adopting a basis
set that is adapted to the molecular geometry. (In contrast, significant
distortions to the electronic structure of atoms and molecules can
be observed \eg{} in strong magnetic fields as discussed in \citeref{Lehtola2019d}
and references therein, in which case LCAO calculations become unreliable.)

Because the electronic structure at $R\to0$ may be quite far from
those for which typical basis sets have been optimized, one can customize
the basis set for the system by hand as in \citeref{Sabelli1979}.
(Alternatively, one could also optimize a new basis set from scratch
for the system.) However, given that this would lead to a different
basis set for every molecule and for every molecular geometry, a systematic
study of the repulsive potentials of all the elements in the periodic
table would be faced with a gargantuan task for basis set generation.
For instance, the PECs for all 4186 diatomic molecules from $Z=1$
to $Z=92$ were calculated in \citeref{Zinoviev2017} at internuclear
distances ranging from $R=0.002$ Å to $R=1000$ Å; this is only feasible
with a fully automatic approach. (Convergence to the basis set limit
was not checked in \citeref{Zinoviev2017}, and we will show later
in the present work that the values are not converged.)

Issues with linear dependencies are encountered also in other applications
of quantum chemistry. For instance, an accurate description of weakly
bound anions may require the use of several shells of diffuse functions
on each atom, making the molecular basis set ill-behaved due to linear
dependencies \citep{Herbert2015}. An approach for curing overcompleteness
issues in the study of weakly bound anions with LCAO basis sets by
pivoted Cholesky decompositions was recently proposed in \citeref{Lehtola2019f}.
Cholesky decomposition algorithms have a long history in quantum chemistry,
starting with with the decomposition of the two-electron integral
tensor proposed over 40 years ago by Beebe and Linderberg \citep{Beebe1977}
that has recently become popular due to efficient implementations
afforded by modern computer architectures, see \eg{} \citep{Folkestad2019,Feng2019}.
The tractability of the Cholesky decomposition of the two-electron
integrals relies on the set of basis function products $\chi_{i}(\boldsymbol{r})\chi_{j}(\boldsymbol{r})$
being highly linearly dependent: due to these dependencies, the number
of Cholesky vectors of the two-electron integrals tensor grows only
linearly with system size. Also method-specific variants related to
the two-electron integrals tensor decomposition have been suggested,
see \citeref{Aquilante2011} for a review. The construction of localized
orbitals by the Cholesky decomposition of the LCAO one-electron density
matrix has also been proposed \citep{Aquilante2006}.

The mathematical closeness of the Cholesky decomposition of the two-electron
integrals tensor, $(ij|kl)=\sum_{P}L_{(ij)}^{P}L_{(kl)}^{P}$ in chemists'
notation, to resolution-of-the-identity methods \citep{Vahtras1993},
$(ij|kl)\approx\sum_{AB}(ij|A)(A|B)^{-1}(B|kl)$, led to a black-box
procedure for the formation of auxiliary basis sets \citep{Aquilante2007a,Aquilante2009},
in which the pivot index of the Cholesky decomposition of the two-electron
integrals is used to determine the auxiliary functions; the resulting
set of auxiliary functions is better behaved than the set of basis
function products. Building on the work of \citep{Aquilante2007a,Aquilante2009},
we proposed curing significant linear dependencies in overcomplete
basis sets by the extraction of a well-behaved subset of the basis
functions using a pivoted Cholesky procedure on the overlap matrix
$S_{\mu\nu}=\langle\mu|\nu\rangle$ to a predefined threshold $\tau$
\citeref{Lehtola2019f}; this yields an optimal approximation for
the original over-complete $\boldsymbol{S}$ \citep{Harbrecht2012}.
Electronic structure calculations can be carried out in the basis
set defined by the set of pivot functions without problems. Before
\citeref{Lehtola2019f}, Cholesky decompositions of the overlap matrix
appear to have only been used in the full, unpivoted form \citep{Millam1997,Challacombe1999,Shao2003}
that is not safe for ill-conditioned matrices $\boldsymbol{S}$.

In the present work, we show that the partial Cholesky decomposition
algorithm proposed in \citeref{Lehtola2019f} for calculations with
basis sets containing a large number of linearly dependent diffuse
functions presents a solution to the present problem of the calculation
of strongly repulsive interatomic potentials by allowing the use of
standard atomic basis sets even at $R\to0$, since the basis function
degeneracies that would otherwise prevent reliable electronic structure
calculations from taking place are cleaned away automatically.

As our aim is simply to prove that the basis set limit can be reached
without problem even at tiny values of $R$, we have chosen to study
a set of seven nuclear reactions involving only closed-shell atoms:
\ce{He2 <=> Be}, \ce{HeNe <=> Mg}, \ce{Ne2 <=> Ca}, \ce{HeAr <=> Ca},
\ce{MgAr <=> Zn}, \ce{Ar2 <=> Kr}, and \ce{NeCa <=> Zn}. We show
that the suggested Cholesky procedure reproduces fully numerical HF
reference values for the reactions, while the values reported in \citeref{Zinoviev2017}
are not converged for small $R$. The present case is much more challenging
than that of \citeref{Lehtola2019f}, as \eg{} in \ce{He2}, \ce{Ne2},
and \ce{Ar2} \emph{all basis functions} of an atom become fully degenerate
with the basis function of the other atom for $R\to0$. Our calculations
will be described in \secref{Computational-Details} and our results
reported in \secref{Results}. The work is briefly summarized and
discussed in \secref{Summary-and-Discussion}. Atomic units are used
throughout the manuscript.

\section{Computational Details\label{sec:Computational-Details}}

The PEC for atoms $A$ and $B$ is defined as
\begin{equation}
V_{AB}(R)=E_{\text{tot}}^{A+B}(R)-E_{\text{el}}^{A}-E_{\text{el}}^{B}\label{eq:pot}
\end{equation}
where $E_{\text{tot}}^{A+B}(R)$ is the total energy from the electronic
structure calculation for the nuclei $A$ and $B$ separated by a
distance of $R$, and $E_{\text{el}}^{A}$ and $E_{\text{el}}^{B}$
are the electronic energies of the non-interacting atoms, respectively.
The total energy $E_{\text{tot}}^{A+B}$ can be decomposed into a
sum of the electronic energy $E_{\text{el}}^{A+B}(R)$ and the nuclear
repulsion energy $E_{\text{nuc}}^{A+B}(R)$. Since the electronic
energy of the compound atom $(A+B)$ is finite, $E_{\text{tot}}^{A+B}(R)$
behaves asymptotically as $E_{\text{tot}}^{A+B}(R)\approx E_{\text{nuc}}^{A+B}(R)=Z_{A}Z_{B}R^{-1}$
for small $R$. Because $V_{AB}(R)$ thus diverges for small $R$,
it is typical to report the PEC in terms of a screening function
\begin{equation}
\Phi_{AB}(R)=\frac{V_{AB}(R)}{E_{\text{nuc}}^{A+B}(R)}=\frac{RV_{AB}(R)}{Z_{A}Z_{B}}\label{eq:screen}
\end{equation}
as it is more easily manipulable, having the limits $\Phi_{AB}(0)=1$
and $\Phi_{AB}(\infty)=0$ .

Although the procedure of \citeref{Lehtola2019f} can be used with
any type of atomic basis set (see \citeref{Lehtola2019c} for a review
thereof), Gaussian basis sets are employed in the present work. Furthermore,
while the approach of \citeref{Lehtola2019f} can also be applied
to density functional or post-HF calculations, the HF level of theory
is used in the present work as it has been found to be sufficient
for the reproduction of repulsive potentials \citep{Nordlund1997}.

The \Erkale{} program \citep{Lehtola2012,erkale} is used for the
Gaussian-basis calculations. The nuclei $A$ and $B$ are placed in
the \Erkale{} calculations along the $z$ axis at $(0,0,-R/2)$ and
$(0,0,R/2)$, respectively, along with their atomic basis functions.
Next, in order to be able to describe the compound atom ($A+B$) limit,
basis functions for the compound atom are included in the calculation;
placing the compound nucleus at the center of charge at $(0,0,{\displaystyle (Z_{B}-Z_{A})R/[2(Z_{B}+Z_{A})]})$
leads to a vanishing dipole moment of the nuclear charge distribution,
and hopefully a more accurate calculation. Once the basis functions
for the compound nucleus have been added, the construction of the
one-electron basis $\{\ket{\mu}\}$ is complete; however, by this
stage the basis set is likely overcomplete.

Next, the overlap matrix $S_{\mu\nu}=\langle\mu|\nu\rangle$, its
eigenvalues $\lambda_{i}$ and its reciprocal condition number
\begin{equation}
r=\frac{\lambda_{\text{min}}}{\lambda_{\text{max}}}\label{eq:rcond}
\end{equation}
 are computed. If the basis set is found to be overcomplete, \ie{},
$r$ is found to be smaller than the machine epsilon, the Cholesky
procedure of \citeref{Lehtola2019f} is used to regularize the molecular
basis set. The procedure uses a pivoted Cholesky decomposition to
pick a subset of the basis functions $\{\ket{\mu}\}$ that spans all
of the functions in the original basis set up to a predefined threshold.
The resulting reduced-size basis is numerically well-conditioned,
and poses no problems to electronic structure calculations which then
proceed as usual. The procedure is implemented as a modification \citep{Lehtola2019f}
to the canonical orthogonalization method \citep{Lowdin1956}; a Cholesky
threshold of $10^{-7}$ and a linear dependence threshold of $10^{-5}$
are used in the present work. As the basis set is normalized, $S_{\mu\mu}=1$,
the first function of the Cholesky procedure will be the first function
in the basis set. However, since diffuse basis functions may be representable
as superpositions of more localized basis functions while the converse
is unlikely to be the case, the basis functions are reorganized from
tight to diffuse before the Cholesky procedure. This procedure was
found to yield more compact reduced basis sets in \citeref{Lehtola2019f}.

The screening function $\Phi(R)$ is computed with \Erkale{} on a
logarithmic grid consisting of 121 points ranging from $R=10^{-5}$
Å to $R=10$ Å. The Gaussian-basis values are then compared to a set
of fully numerical reference values obtained with the \HelFEM{} program
\citep{Lehtola2019a,Lehtola2019b,HelFEM}. The superposition of atomic
potentials (SAP) initial guess \citep{Lehtola2019} is used in all
\Erkale{} and \HelFEM{} calculations in combination with local exchange
potentials recently determined at the complete basis set limit \citep{Lehtola2019e}.
The SAP guess correctly includes the significant Pauli repulsion between
the electrons on the two nuclei at small $R$ in contrast to its commonly-used
alternatives discussed in \citeref{Lehtola2019}, thus leading to
faster convergence of the self-consistent field procedure.

Only singlet $\Sigma$ wave functions are considered in the present
work, in analogy to \citeref{Nordlund1997}. In the cases of \ce{He2 <=> Be},
\ce{HeNe <=> Mg}, \ce{Ne2 <=> Ca}, and \ce{HeAr <=> Ca}, the large-$R$
and small-$R$ wave functions have the same electronic configurations:
two occupied $\sigma$ orbitals for \ce{He2} and \ce{Be}, four $\sigma$
and one $\pi$ orbital for \ce{HeNe} and \ce{Mg}, and six $\sigma$
and two $\pi$ orbitals for \ce{Ne2}, \ce{Ca}, and \ce{HeAr}; each
$\sigma$ and $\pi$ orbital fitting two and four electrons, respectively
\citep{Lehtola2019c}. For the heavier systems, \ce{MgAr <=> Zn},
\ce{Ar2 <=> Kr}, and \ce{NeCa <=> Zn}, the electronic configurations
are different at small $R$ and at large $R$, and both states were
calculated: nine $\sigma$ and three $\pi$ in \ce{MgAr} and \ce{NeCa};
seven $\sigma$, three $\pi$ and one $\delta$ orbital in Zn; ten
$\sigma$ and four $\pi$ orbitals in \ce{Ar2}; and eight $\sigma$,
four $\pi$ and one $\delta$ orbital in Kr; $\delta$ orbitals likewise
fitting four electrons \citep{Lehtola2019c}. The values reported
correspond to the lower state in each case; for instance, the Kr configuration
is lower in \ce{Ar2} for $R\lesssim0.56$ Å, the state crossing depending
on the used basis set.

\section{Results\label{sec:Results}}

Very accurate LCAO calculations can be performed both at small $R$
and at large $R$, as in the former case a single expansion center
is sufficient, and as in the latter the basis functions on the two
centers do not develop strong linear dependencies. For this reason,
we start off in \tabref{screening} by comparing the values of the
screening function $\Phi(R)$ at intermediate values of $R$ for the
decontracted double- to quadruple-$\zeta$ pc-$n$ basis sets \citep{Jensen2001}
(denoted as un-pc-1, un-pc-2, and un-pc-3, respectively) as well as
for the universal Gaussian basis set \citep{DeCastro1998} (UGBS)
to fully numerical reference values.

Examination of the data in \tabref{screening} shows that good results
are already obtained with the double-$\zeta$ un-pc-1 basis set, while
the UGBS basis set appears to reproduce values that are in-between
those of the triple-$\zeta$ un-pc-2 and the quadruple-$\zeta$ un-pc-3
basis set at small $R$. This suggests that the screening function
is insensitive to polarization functions at small $R$; however, some
polarization effects are already described by the compound nucleus
basis functions included at the center of charge. As the UGBS basis
set is available for most of the periodic table and equivalent atomic
basis sets can be easily generated, see \citeref{Lehtola2020b}, we
choose the UGBS basis set for the rest of the work.

To confirm the finding of \citeref{Nordlund1997} that the screening
function has a negligible dependence on the employed level of theory,
we also report fully numerical reference values for \ce{Ar2} calculated
with \HelFEM{} using the local density approximation (LDA), in which
the local exchange functional \citep{Bloch1929,Dirac1930} is combined
with the Vosko--Wilk--Nusair correlation functional \citep{Vosko1980}
as in \citeref{Nordlund1997}. The differences of the HF and LDA screening
functions are only seen in the third decimal, confirming that HF or
DFT is suitable for the present purposes.

\begin{figure}
\begin{centering}
\includegraphics[width=0.45\textwidth]{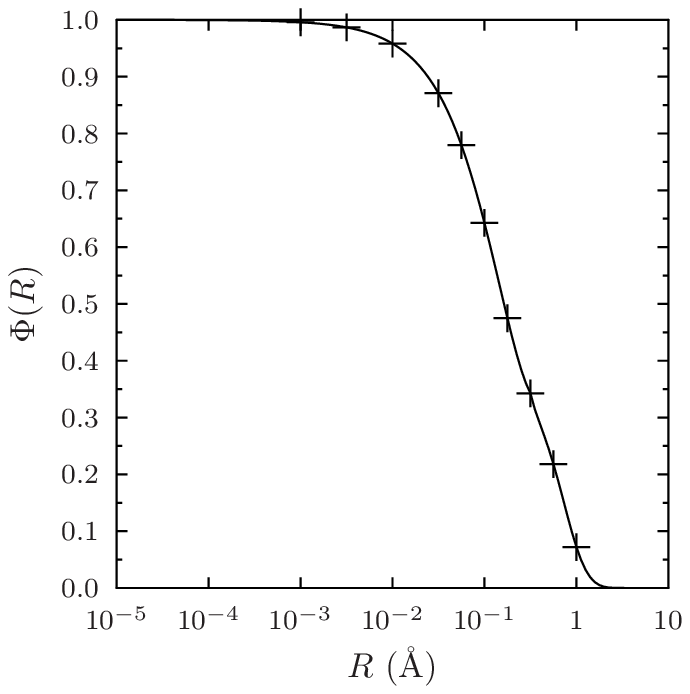}
\par\end{centering}
\caption{UGBS screening function for the \ce{He2 <=> Be} reaction with fully
numerical reference values (+).\label{fig:He2}}
\end{figure}

\begin{figure}
\begin{centering}
\includegraphics[width=0.45\textwidth]{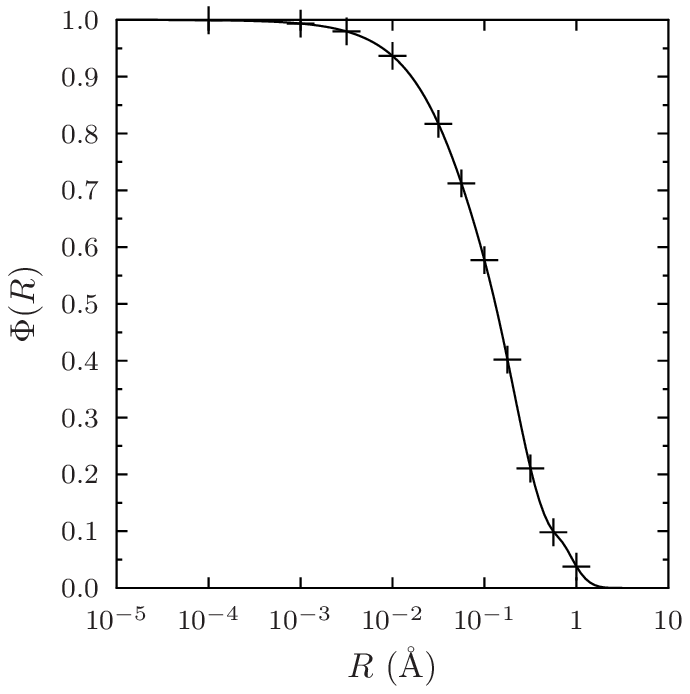}
\par\end{centering}
\caption{UGBS screening function for the \ce{HeNe <=> Mg} reaction with fully
numerical reference values (+).\label{fig:HeNe}}
\end{figure}

\begin{figure}
\begin{centering}
\includegraphics[width=0.45\textwidth]{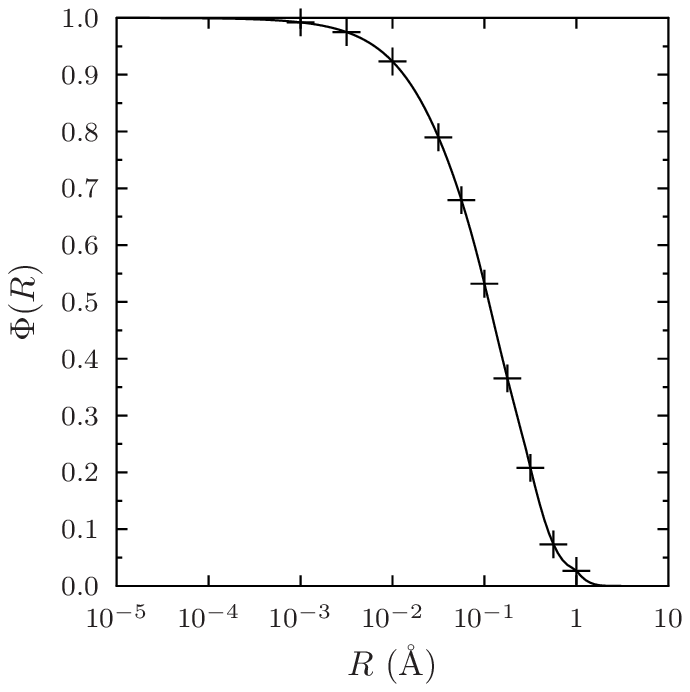}
\par\end{centering}
\caption{UGBS screening function for the \ce{Ne2 <=> Ca} reaction with fully
numerical reference values (+).\label{fig:Ne2}}
\end{figure}

\begin{figure}
\begin{centering}
\includegraphics[width=0.45\textwidth]{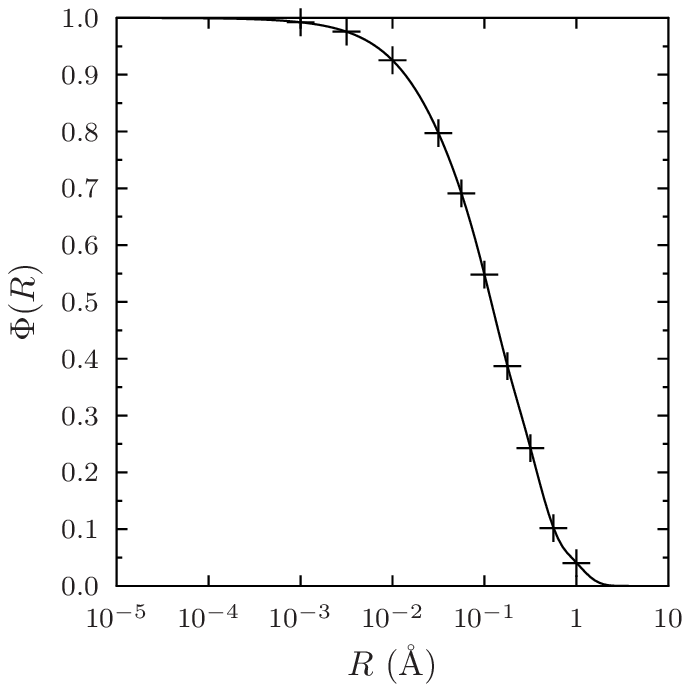}
\par\end{centering}
\caption{UGBS screening function for the \ce{HeAr <=> Ca} reaction with fully
numerical reference values (+).\label{fig:HeAr}}
\end{figure}

\begin{figure}
\begin{centering}
\includegraphics[width=0.45\textwidth]{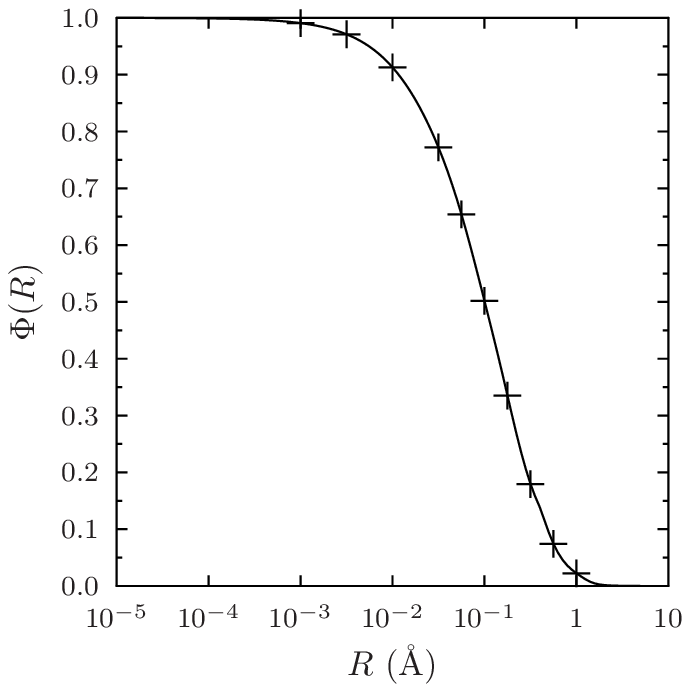}
\par\end{centering}
\caption{UGBS screening function for the \ce{MgAr <=> Zn} reaction with fully
numerical reference values (+).\label{fig:MgAr}}
\end{figure}

\begin{figure}
\begin{centering}
\includegraphics[width=0.45\textwidth]{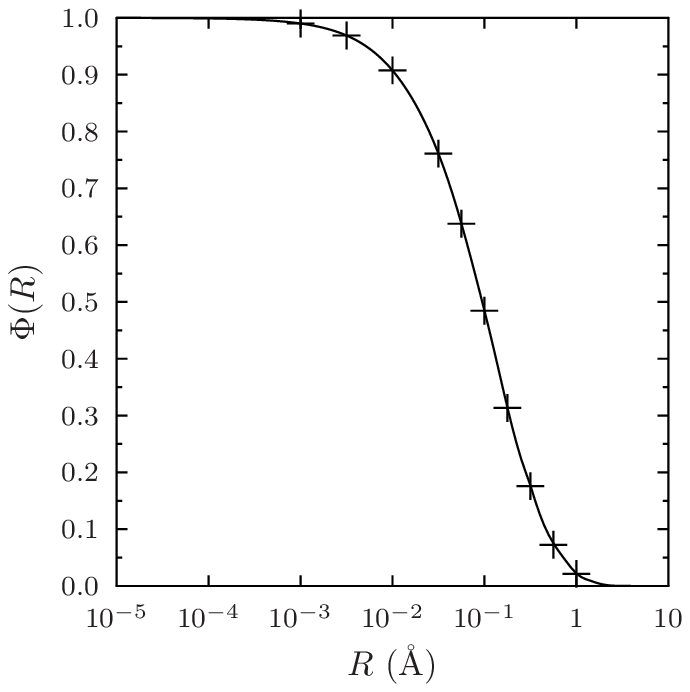}
\par\end{centering}
\caption{UGBS screening function for the \ce{ArAr <=> Kr} reaction with fully
numerical reference values (+).\label{fig:Ar2}}
\end{figure}

\begin{figure}
\begin{centering}
\includegraphics[width=0.45\textwidth]{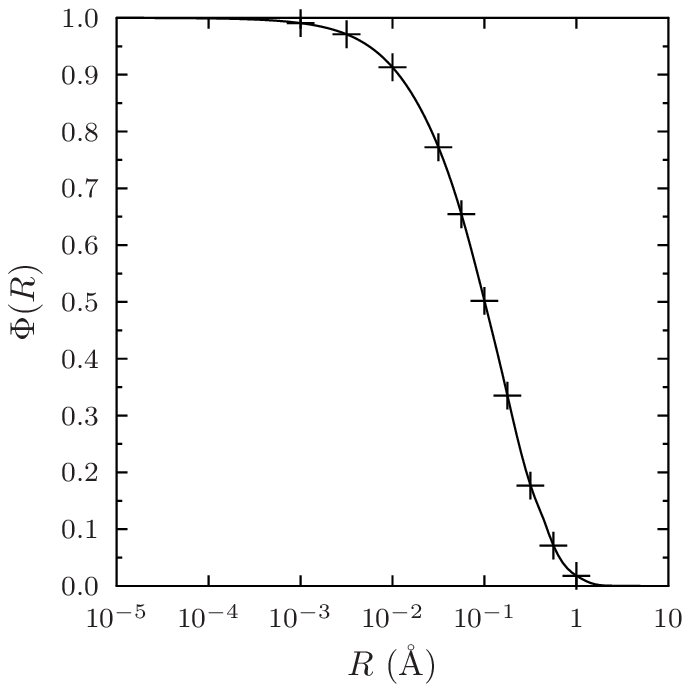}
\par\end{centering}
\caption{UGBS screening function for the \ce{NeCa <=> Zn} reaction with fully
numerical reference values (+).\label{fig:CaNe}}
\end{figure}

The screening functions for the seven nuclear reactions computed with
the UGBS basis set are shown in \figref{He2} for \ce{He2 <=> Be},
\figref{HeNe} for \ce{HeNe <=> Mg}, \figref{Ne2} for \ce{Ne2 <=> Ca},
\figref{HeAr} for \ce{HeAr <=> Ca}, \figref{MgAr} for \ce{MgAr <=> Zn},
\figref{Ar2} for \ce{Ar2 <=> Kr}, and \figref{CaNe} for \ce{NeCa <=> Zn}.
The curves are smooth and the agreement with fully numerical reference
values is superb in all cases. 

\begin{figure}
\begin{centering}
\includegraphics[width=0.45\textwidth]{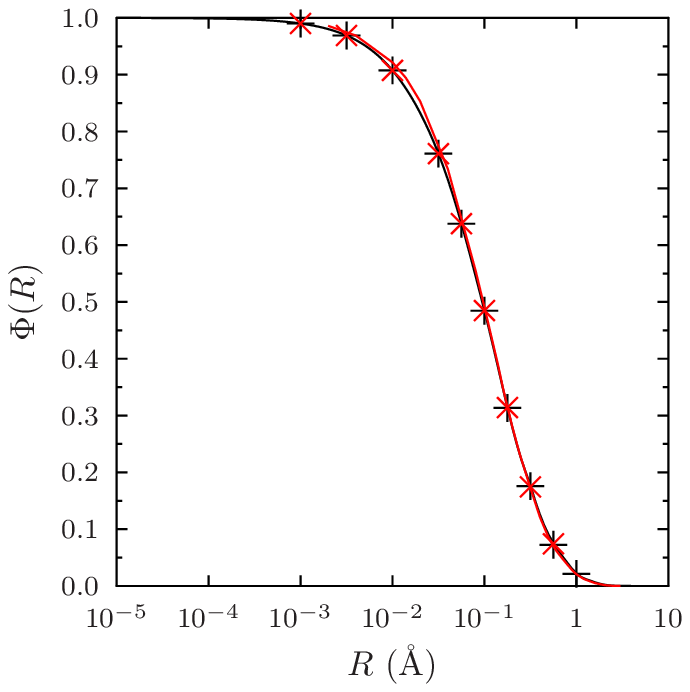}
\par\end{centering}
\caption{Comparison of the \ce{ArAr <=> Kr} UGBS/Hartree--Fock screening
function (black line) against LDA data from \citeref{Zinoviev2017}
(red line). Fully numerical Hartree--Fock (black +) as well as LDA
(red $\times$) reference values are also shown.\label{fig:zino}}
\end{figure}

All of these reactions have also been studied in \citeref{Zinoviev2017}
with the LDA approach of \citeref{Nordlund1997} \bibnote{Kai Nordlund, private communication, 2020.}.
However, out of the seven reactions currently examined, \citeref{Zinoviev2017}
only only reports data for \ce{Ar2 <=> Kr}. A comparison to the UGBS
results and fully numerical HF and LDA reference values is shown in
\figref{zino}. The data from \citeref{Zinoviev2017} agree with the
present values at large $R$, but discrepancies are visible for $R<0.1$
Å. The UGBS data is agrees with the fully numerical HF and LDA reference
data, indicating that an insufficient basis set was used in \citeref{Zinoviev2017}.

\begin{table*}
\begin{centering}
\begin{tabular}{lccccccccccc}
 &  & $10^{-3}$ & $10^{-2.5}$ & $10^{-2}$ & $10^{-1.5}$ & $10^{-1.25}$ & $10^{-1}$ & $10^{-0.75}$ & $10^{-0.5}$ & $10^{-0.25}$ & $10^{0}$\tabularnewline
\hline 
\hline 
\ce{He2} & \HelFEM{} & 0.99582 & 0.98678 & 0.95831 & 0.87089 & 0.77943 & 0.64265 & 0.47484 & 0.34264 & 0.21805 & 0.07205\tabularnewline
 & $\Delta$un-pc-1 & 0.00000 & -0.00001 & -0.00005 & -0.00007 & 0.00027 & 0.00095 & 0.00068 & 0.00023 & 0.00162 & 0.00011\tabularnewline
 & $\Delta$un-pc-2 & 0.00000 & 0.00000 & 0.00000 & 0.00004 & 0.00006 & 0.00008 & 0.00007 & 0.00005 & 0.00041 & -0.00005\tabularnewline
 & $\Delta$un-pc-3 & 0.00000 & 0.00000 & 0.00000 & 0.00000 & 0.00000 & 0.00000 & 0.00001 & 0.00002 & 0.00002 & 0.00000\tabularnewline
 & $\Delta$UGBS & 0.00000 & 0.00000 & 0.00000 & 0.00000 & 0.00000 & 0.00000 & 0.00001 & 0.00006 & 0.00021 & 0.00022\tabularnewline
 &  &  &  &  &  &  &  &  &  &  & \tabularnewline
HeNe & \HelFEM{} & 0.99356 & 0.97966 & 0.93658 & 0.81686 & 0.71223 & 0.57716 & 0.40213 & 0.21037 & 0.09810 & 0.03762\tabularnewline
 & $\Delta$un-pc-1 & -0.00001 & -0.00003 & -0.00007 & 0.00000 & 0.00013 & -0.00007 & -0.00004 & 0.00016 & -0.00022 & -0.00111\tabularnewline
 & $\Delta$un-pc-2 & 0.00000 & 0.00000 & 0.00000 & 0.00006 & 0.00006 & 0.00009 & 0.00024 & 0.00037 & 0.00014 & -0.00010\tabularnewline
 & $\Delta$un-pc-3 & 0.00000 & 0.00000 & 0.00000 & 0.00000 & 0.00001 & 0.00004 & 0.00006 & 0.00003 & 0.00002 & 0.00000\tabularnewline
 & $\Delta$UGBS & 0.00000 & 0.00000 & 0.00000 & 0.00000 & 0.00001 & 0.00008 & 0.00031 & 0.00085 & 0.00120 & 0.00031\tabularnewline
 &  &  &  &  &  &  &  &  &  &  & \tabularnewline
\ce{Ne2} & \HelFEM{} & 0.99207 & 0.97503 & 0.92326 & 0.78984 & 0.67932 & 0.53204 & 0.36532 & 0.20790 & 0.07321 & 0.02656\tabularnewline
 & $\Delta$un-pc-1 & 0.00000 & -0.00001 & 0.00001 & -0.00001 & 0.00008 & 0.00053 & 0.00065 & 0.00029 & -0.00007 & -0.00035\tabularnewline
 & $\Delta$un-pc-2 & 0.00000 & 0.00000 & 0.00000 & 0.00001 & 0.00006 & 0.00009 & 0.00007 & 0.00009 & 0.00007 & -0.00001\tabularnewline
 & $\Delta$un-pc-3 & 0.00000 & 0.00000 & 0.00000 & 0.00000 & 0.00001 & 0.00001 & 0.00002 & 0.00003 & 0.00001 & 0.00001\tabularnewline
 & $\Delta$UGBS & 0.00000 & 0.00000 & 0.00000 & 0.00000 & 0.00000 & 0.00001 & 0.00002 & 0.00012 & 0.00020 & 0.00011\tabularnewline
 &  &  &  &  &  &  &  &  &  &  & \tabularnewline
HeAr & \HelFEM{} & 0.99228 & 0.97569 & 0.92539 & 0.79716 & 0.69117 & 0.54811 & 0.38703 & 0.24271 & 0.10196 & 0.04028\tabularnewline
 & $\Delta$un-pc-1 & 0.00000 & 0.00000 & 0.00001 & 0.00012 & 0.00029 & 0.00058 & 0.00060 & 0.00043 & 0.00032 & 0.00001\tabularnewline
 & $\Delta$un-pc-2 & 0.00000 & 0.00000 & 0.00001 & 0.00008 & 0.00008 & 0.00019 & 0.00028 & 0.00020 & 0.00009 & 0.00001\tabularnewline
 & $\Delta$un-pc-3 & 0.00000 & 0.00000 & 0.00000 & 0.00001 & 0.00004 & 0.00010 & 0.00011 & 0.00006 & 0.00002 & 0.00001\tabularnewline
 & $\Delta$UGBS & 0.00000 & 0.00000 & 0.00000 & 0.00001 & 0.00004 & 0.00016 & 0.00038 & 0.00043 & 0.00084 & 0.00083\tabularnewline
 &  &  &  &  &  &  &  &  &  &  & \tabularnewline
MgAr & \HelFEM{} & 0.99081 & 0.97114 & 0.91299 & 0.77196 & 0.65409 & 0.50202 & 0.33521 & 0.17926 & 0.07422 & 0.02229\tabularnewline
 & $\Delta$un-pc-1 & 0.00000 & 0.00000 & 0.00000 & 0.00003 & 0.00006 & 0.00008 & 0.00024 & 0.00091 & 0.00032 & 0.00018\tabularnewline
 & $\Delta$un-pc-2 & 0.00000 & 0.00000 & 0.00000 & 0.00001 & 0.00002 & 0.00004 & 0.00017 & 0.00056 & 0.00012 & 0.00009\tabularnewline
 & $\Delta$un-pc-3 & 0.00000 & 0.00000 & 0.00000 & 0.00000 & 0.00000 & 0.00002 & 0.00009 & 0.00018 & 0.00005 & 0.00002\tabularnewline
 & $\Delta$UGBS & 0.00000 & 0.00000 & 0.00000 & 0.00000 & 0.00001 & 0.00003 & 0.00020 & 0.00086 & 0.00029 & 0.00025\tabularnewline
 &  &  &  &  &  &  &  &  &  &  & \tabularnewline
\ce{Ar2} & \HelFEM{} & 0.99011 & 0.96900 & 0.90749 & 0.76090 & 0.63761 & 0.48451 & 0.31334 & 0.17568 & 0.07255 & 0.02137\tabularnewline
 & $\Delta$LDA$^{a}$ & 0.00001 & 0.00003 & 0.00009 & 0.00012 & 0.00017 & 0.00028 & 0.00073 & -0.00130 & 0.00174 & no data\tabularnewline
 & $\Delta$un-pc-1 & 0.00000 & 0.00000 & 0.00000 & 0.00003 & 0.00005 & 0.00013 & 0.00089 & 0.00158 & 0.00290 & 0.00155\tabularnewline
 & $\Delta$un-pc-2 & 0.00000 & 0.00000 & 0.00000 & 0.00001 & 0.00002 & 0.00009 & 0.00080 & 0.00117 & 0.00115 & 0.00066\tabularnewline
 & $\Delta$un-pc-3 & 0.00000 & 0.00000 & 0.00000 & 0.00000 & 0.00001 & 0.00006 & 0.00055 & 0.00029 & 0.00010 & 0.00030\tabularnewline
 & $\Delta$UGBS & 0.00000 & 0.00000 & 0.00000 & 0.00000 & 0.00001 & 0.00009 & 0.00078 & 0.00141 & 0.00275 & 0.00078\tabularnewline
 &  &  &  &  &  &  &  &  &  &  & \tabularnewline
NeCa & \HelFEM{} & 0.99082 & 0.97117 & 0.91310 & 0.77230 & 0.65438 & 0.50203 & 0.33509 & 0.17662 & 0.07093 & 0.01768\tabularnewline
 & $\Delta$un-pc-1 & 0.00000 & 0.00000 & 0.00002 & 0.00005 & 0.00012 & 0.00032 & 0.00057 & 0.00061 & 0.00023 & 0.00008\tabularnewline
 & $\Delta$un-pc-2 & 0.00000 & 0.00000 & 0.00000 & 0.00002 & 0.00005 & 0.00007 & 0.00016 & 0.00026 & 0.00011 & 0.00004\tabularnewline
 & $\Delta$un-pc-3 & 0.00000 & 0.00000 & 0.00000 & 0.00000 & 0.00001 & 0.00002 & 0.00006 & 0.00007 & 0.00002 & 0.00001\tabularnewline
 & $\Delta$UGBS & 0.00000 & 0.00000 & 0.00000 & 0.00000 & 0.00001 & 0.00003 & 0.00021 & 0.00095 & 0.00047 & 0.00049\tabularnewline
\end{tabular}
\par\end{centering}
\caption{Values of screening function $\Phi(R)$ computed at various points
$R$ (value in Å given on the first row) with the fully numerical
\HelFEM{} program. The Gaussian-basis-set truncation errors $\Delta\text{basis}=\Phi^{\text{basis}}(R)-\Phi^{\text{reference}}(R)$
of the un-pc-$n$ and UGBS basis sets are also shown; these calculations
were done with \Erkale{}. The data for \ce{Ar2} also includes the
differences between the LDA and HF screening functions' reference
values $\Delta\text{LDA}=\Phi^{\text{LDA}}(R)-\Phi^{\text{HF}}(R)$,
both of which have been computed with \HelFEM{}.\label{tab:screening}}
\end{table*}


\section{Summary and Discussion\label{sec:Summary-and-Discussion}}

We have shown by comparison to fully numerical Hartree--Fock reference
values that accurate potential energy curves can be reproduced with
linear combination of atomic orbitals (LCAO) calculations even in
the strongly repulsive region at small internuclear distances---where
even the core orbital basis functions become fully linearly dependent---by
using a recently suggested procedure \citep{Lehtola2019f} to eliminate
linear dependencies from the basis set. As LCAO calculations are faster
and easier to run than fully numerical ones, the automated procedure
of the present work enables the systematical calculation of screening
functions along the lines of \citeref{Zinoviev2017} but with guaranteed
accuracy. 

Although the present study has been limited to diatomic molecules,
the underlying Cholesky decomposition has already been shown to work
for polyatomic molecules \citep{Lehtola2019f}, making the calculation
of repulsive many-body potentials also possible.

The facile computation of the repulsive barrier afforded by the present
method should make it easier to study various irradiation processes,
in which the purely repulsive part of the potential plays a pivotal
role. For instance, defect formation and migration in materials subjected
to particle bombardment is determined purely by the repulsive part
of the potential \citep{Juslin2008}, and accounting for this kind
of radiation damage is an important aspect in the design of radiation
shielding materials of fusion reactors \citep{Becquart2011,Sand2014}.

The present study has been limited to non-relativistic calculations
on light, closed-shell atoms. As relativistic effects increase rapidly
in $Z$ \citep{Pyykko2012,Pyykko2012a}, they are more important at
the compound nucleus limit $R\to0$ than at large $R$. Note also
that in contrast to usual applications to chemistry, the screening
function merits from no systematic error cancellation from the subtraction
of atomic energies. The present procedure can, however, be straightforwardly
extended to relativistic methods as well, making it possible to model
the relativistic effects. Open-shell atoms as well as relativistic
effects will be visited in future work.

\section*{Acknowledgments}

I thank Kai Nordlund and Dage Sundholm for discussions on repulsive
potentials. This work has been supported by the Academy of Finland
(Suomen Akatemia) through project number 311149. Computational resources
provided by CSC -- It Center for Science Ltd (Espoo, Finland) and
the Finnish Grid and Cloud Infrastructure (persistent identifier urn:nbn:fi:research-infras-2016072533)
are gratefully acknowledged.

\bibliography{citations}

\begin{thebibliography}{52}%
\makeatletter
\providecommand \@ifxundefined [1]{%
 \@ifx{#1\undefined}
}%
\providecommand \@ifnum [1]{%
 \ifnum #1\expandafter \@firstoftwo
 \else \expandafter \@secondoftwo
 \fi
}%
\providecommand \@ifx [1]{%
 \ifx #1\expandafter \@firstoftwo
 \else \expandafter \@secondoftwo
 \fi
}%
\providecommand \natexlab [1]{#1}%
\providecommand \enquote  [1]{``#1''}%
\providecommand \bibnamefont  [1]{#1}%
\providecommand \bibfnamefont [1]{#1}%
\providecommand \citenamefont [1]{#1}%
\providecommand \href@noop [0]{\@secondoftwo}%
\providecommand \href [0]{\begingroup \@sanitize@url \@href}%
\providecommand \@href[1]{\@@startlink{#1}\@@href}%
\providecommand \@@href[1]{\endgroup#1\@@endlink}%
\providecommand \@sanitize@url [0]{\catcode `\\12\catcode `\$12\catcode
  `\&12\catcode `\#12\catcode `\^12\catcode `\_12\catcode `\%12\relax}%
\providecommand \@@startlink[1]{}%
\providecommand \@@endlink[0]{}%
\providecommand \url  [0]{\begingroup\@sanitize@url \@url }%
\providecommand \@url [1]{\endgroup\@href {#1}{\urlprefix }}%
\providecommand \urlprefix  [0]{URL }%
\providecommand \Eprint [0]{\href }%
\providecommand \doibase [0]{http://dx.doi.org/}%
\providecommand \selectlanguage [0]{\@gobble}%
\providecommand \bibinfo  [0]{\@secondoftwo}%
\providecommand \bibfield  [0]{\@secondoftwo}%
\providecommand \translation [1]{[#1]}%
\providecommand \BibitemOpen [0]{}%
\providecommand \bibitemStop [0]{}%
\providecommand \bibitemNoStop [0]{.\EOS\space}%
\providecommand \EOS [0]{\spacefactor3000\relax}%
\providecommand \BibitemShut  [1]{\csname bibitem#1\endcsname}%
\let\auto@bib@innerbib\@empty
%</preamble>
\bibitem [{\citenamefont {Sigmund}(2014)}]{Sigmund2014}%
  \BibitemOpen
  \bibfield  {author} {\bibinfo {author} {\bibfnamefont {P.}~\bibnamefont
  {Sigmund}},\ }\href {\doibase 10.1007/978-3-319-05564-0} {\emph {\bibinfo
  {title} {{Particle Penetration and Radiation Effects Volume 2 -- Penetration
  of Atomic and Molecular Ions}}}},\ \bibinfo {series} {Springer Series in
  Solid-State Sciences}, Vol.\ \bibinfo {volume} {179}\ (\bibinfo  {publisher}
  {Springer Cham Heidelberg New York Dordrecht London},\ \bibinfo {year}
  {2014})\ p.\ \bibinfo {pages} {617}\BibitemShut {NoStop}%
\bibitem [{\citenamefont {Karolewski}(2007)}]{Karolewski2007}%
  \BibitemOpen
  \bibfield  {author} {\bibinfo {author} {\bibfnamefont {M.~A.}\ \bibnamefont
  {Karolewski}},\ }\href {\doibase 10.1016/j.nimb.2006.12.027} {\bibfield
  {journal} {\bibinfo  {journal} {Nucl. Instrum. Methods Phys. Res., Sect. B}\ }\textbf {\bibinfo {volume} {256}},\
  \bibinfo {pages} {354} (\bibinfo {year} {2007})}\BibitemShut {NoStop}%
\bibitem [{\citenamefont {Ziegler}\ \emph {et~al.}(1985)\citenamefont
  {Ziegler}, \citenamefont {Littmark},\ and\ \citenamefont
  {Biersack}}]{Ziegler1985}%
  \BibitemOpen
  \bibfield  {author} {\bibinfo {author} {\bibfnamefont {J.~F.}\ \bibnamefont
  {Ziegler}}, \bibinfo {author} {\bibfnamefont {U.}~\bibnamefont {Littmark}}, \
  and\ \bibinfo {author} {\bibfnamefont {J.~P.}\ \bibnamefont {Biersack}},\
  }\href@noop {} {\emph {\bibinfo {title} {{The stopping and range of ions in
  solids}}}}\ (\bibinfo  {publisher} {Pergamon, New York},\ \bibinfo {year}
  {1985})\BibitemShut {NoStop}%
\bibitem [{\citenamefont {Hohenberg}\ and\ \citenamefont
  {Kohn}(1964)}]{Hohenberg1964}%
  \BibitemOpen
  \bibfield  {author} {\bibinfo {author} {\bibfnamefont {P.}~\bibnamefont
  {Hohenberg}}\ and\ \bibinfo {author} {\bibfnamefont {W.}~\bibnamefont
  {Kohn}},\ }\href {\doibase 10.1103/PhysRev.136.B864} {\bibfield  {journal}
  {\bibinfo  {journal} {Phys. Rev.}\ }\textbf {\bibinfo {volume} {136}},\
  \bibinfo {pages} {B864} (\bibinfo {year} {1964})}\BibitemShut {NoStop}%
\bibitem [{\citenamefont {Kohn}\ and\ \citenamefont {Sham}(1965)}]{Kohn1965}%
  \BibitemOpen
  \bibfield  {author} {\bibinfo {author} {\bibfnamefont {W.}~\bibnamefont
  {Kohn}}\ and\ \bibinfo {author} {\bibfnamefont {L.~J.}\ \bibnamefont
  {Sham}},\ }\href {\doibase 10.1103/PhysRev.140.A1133} {\bibfield  {journal}
  {\bibinfo  {journal} {Phys. Rev.}\ }\textbf {\bibinfo {volume} {140}},\
  \bibinfo {pages} {A1133} (\bibinfo {year} {1965})}\BibitemShut {NoStop}%
\bibitem [{\citenamefont {Sabelli}\ \emph {et~al.}(1978)\citenamefont
  {Sabelli}, \citenamefont {Kantor}, \citenamefont {Benedek},\ and\
  \citenamefont {Gilbert}}]{Sabelli1978}%
  \BibitemOpen
  \bibfield  {author} {\bibinfo {author} {\bibfnamefont {N.~H.}\ \bibnamefont
  {Sabelli}}, \bibinfo {author} {\bibfnamefont {M.}~\bibnamefont {Kantor}},
  \bibinfo {author} {\bibfnamefont {R.}~\bibnamefont {Benedek}}, \ and\
  \bibinfo {author} {\bibfnamefont {T.~L.}\ \bibnamefont {Gilbert}},\ }\href
  {\doibase 10.1063/1.436068} {\bibfield  {journal} {\bibinfo  {journal} {J.
  Chem. Phys.}\ }\textbf {\bibinfo {volume} {68}},\ \bibinfo {pages} {2767}
  (\bibinfo {year} {1978})}\BibitemShut {NoStop}%
\bibitem [{\citenamefont {Sabelli}\ \emph {et~al.}(1979)\citenamefont
  {Sabelli}, \citenamefont {Benedek},\ and\ \citenamefont
  {Gilbert}}]{Sabelli1979}%
  \BibitemOpen
  \bibfield  {author} {\bibinfo {author} {\bibfnamefont {N.~H.}\ \bibnamefont
  {Sabelli}}, \bibinfo {author} {\bibfnamefont {R.}~\bibnamefont {Benedek}}, \
  and\ \bibinfo {author} {\bibfnamefont {T.~L.}\ \bibnamefont {Gilbert}},\
  }\href {\doibase 10.1103/PhysRevA.20.677} {\bibfield  {journal} {\bibinfo
  {journal} {Phys. Rev. A}\ }\textbf {\bibinfo {volume} {20}},\ \bibinfo
  {pages} {677} (\bibinfo {year} {1979})}\BibitemShut {NoStop}%
\bibitem [{\citenamefont {Keinonen}\ \emph {et~al.}(1991)\citenamefont
  {Keinonen}, \citenamefont {Kuronen}, \citenamefont {Tikkanen}, \citenamefont
  {B{\"{o}}rner}, \citenamefont {Jolie}, \citenamefont {Ulbig}, \citenamefont
  {Kessler}, \citenamefont {Nieminen}, \citenamefont {Puska},\ and\
  \citenamefont {Seitsonen}}]{Keinonen1991}%
  \BibitemOpen
  \bibfield  {author} {\bibinfo {author} {\bibfnamefont {J.}~\bibnamefont
  {Keinonen}}, \bibinfo {author} {\bibfnamefont {A.}~\bibnamefont {Kuronen}},
  \bibinfo {author} {\bibfnamefont {P.}~\bibnamefont {Tikkanen}}, \bibinfo
  {author} {\bibfnamefont {H.~G.}\ \bibnamefont {B{\"{o}}rner}}, \bibinfo
  {author} {\bibfnamefont {J.}~\bibnamefont {Jolie}}, \bibinfo {author}
  {\bibfnamefont {S.}~\bibnamefont {Ulbig}}, \bibinfo {author} {\bibfnamefont
  {E.~G.}\ \bibnamefont {Kessler}}, \bibinfo {author} {\bibfnamefont {R.~M.}\
  \bibnamefont {Nieminen}}, \bibinfo {author} {\bibfnamefont {M.~J.}\
  \bibnamefont {Puska}}, \ and\ \bibinfo {author} {\bibfnamefont {A.~P.}\
  \bibnamefont {Seitsonen}},\ }\href {\doibase 10.1103/PhysRevLett.67.3692}
  {\bibfield  {journal} {\bibinfo  {journal} {Phys. Rev. Lett.}\ }\textbf
  {\bibinfo {volume} {67}},\ \bibinfo {pages} {3692} (\bibinfo {year}
  {1991})}\BibitemShut {NoStop}%
\bibitem [{\citenamefont {Keinonen}\ \emph {et~al.}(1994)\citenamefont
  {Keinonen}, \citenamefont {Kuronen}, \citenamefont {Nordlund}, \citenamefont
  {Nieminen},\ and\ \citenamefont {Seitsonen}}]{Keinonen1994}%
  \BibitemOpen
  \bibfield  {author} {\bibinfo {author} {\bibfnamefont {J.}~\bibnamefont
  {Keinonen}}, \bibinfo {author} {\bibfnamefont {A.}~\bibnamefont {Kuronen}},
  \bibinfo {author} {\bibfnamefont {K.}~\bibnamefont {Nordlund}}, \bibinfo
  {author} {\bibfnamefont {R.~M.}\ \bibnamefont {Nieminen}}, \ and\ \bibinfo
  {author} {\bibfnamefont {A.~P.}\ \bibnamefont {Seitsonen}},\ }\href {\doibase
  10.1016/0168-583X(94)95387-2} {\bibfield  {journal} {\bibinfo  {journal}
  {Nucl. Instrum. Methods Phys. Res., Sect. B}\ }\textbf {\bibinfo {volume} {88}},\ \bibinfo {pages} {382} (\bibinfo
  {year} {1994})}\BibitemShut {NoStop}%
\bibitem [{\citenamefont {Nordlund}\ \emph {et~al.}(1997)\citenamefont
  {Nordlund}, \citenamefont {Runeberg},\ and\ \citenamefont
  {Sundholm}}]{Nordlund1997}%
  \BibitemOpen
  \bibfield  {author} {\bibinfo {author} {\bibfnamefont {K.}~\bibnamefont
  {Nordlund}}, \bibinfo {author} {\bibfnamefont {N.}~\bibnamefont {Runeberg}},
  \ and\ \bibinfo {author} {\bibfnamefont {D.}~\bibnamefont {Sundholm}},\
  }\href {\doibase 10.1016/S0168-583X(97)00447-3} {\bibfield  {journal}
  {\bibinfo  {journal} {Nucl. Instrum. Methods Phys. Res., Sect. B}\ }\textbf {\bibinfo {volume} {132}},\ \bibinfo
  {pages} {45} (\bibinfo {year} {1997})}\BibitemShut {NoStop}%
\bibitem [{\citenamefont {Pruneda}\ and\ \citenamefont
  {Artacho}(2004)}]{Pruneda2004}%
  \BibitemOpen
  \bibfield  {author} {\bibinfo {author} {\bibfnamefont {J.~M.}\ \bibnamefont
  {Pruneda}}\ and\ \bibinfo {author} {\bibfnamefont {E.}~\bibnamefont
  {Artacho}},\ }\href {\doibase 10.1103/PhysRevB.70.035106} {\bibfield
  {journal} {\bibinfo  {journal} {Phys. Rev. B}\ }\textbf {\bibinfo {volume}
  {70}},\ \bibinfo {pages} {035106} (\bibinfo {year} {2004})},\ \Eprint
  {http://arxiv.org/abs/0401265} {arXiv:0401265 [cond-mat]} \BibitemShut
  {NoStop}%
\bibitem [{\citenamefont {Kuzmin}(2006)}]{Kuzmin2006}%
  \BibitemOpen
  \bibfield  {author} {\bibinfo {author} {\bibfnamefont {V.}~\bibnamefont
  {Kuzmin}},\ }\href {\doibase 10.1016/j.nimb.2006.03.012} {\bibfield
  {journal} {\bibinfo  {journal} {Nucl. Instrum. Methods Phys. Res., Sect. B}\ }\textbf {\bibinfo {volume} {249}},\
  \bibinfo {pages} {13} (\bibinfo {year} {2006})}\BibitemShut {NoStop}%
\bibitem [{\citenamefont {Karolewski}(2006)}]{Karolewski2006}%
  \BibitemOpen
  \bibfield  {author} {\bibinfo {author} {\bibfnamefont {M.~A.}\ \bibnamefont
  {Karolewski}},\ }\href {\doibase 10.1016/j.nimb.2005.07.192} {\bibfield
  {journal} {\bibinfo  {journal} {Nucl. Instrum. Methods Phys. Res., Sect. B}\ }\textbf {\bibinfo {volume} {243}},\
  \bibinfo {pages} {43} (\bibinfo {year} {2006})}\BibitemShut {NoStop}%
\bibitem [{\citenamefont {Kuzmin}(2007)}]{Kuzmin2007}%
  \BibitemOpen
  \bibfield  {author} {\bibinfo {author} {\bibfnamefont {V.}~\bibnamefont
  {Kuzmin}},\ }\href {\doibase 10.1016/j.surfcoat.2006.10.053} {\bibfield
  {journal} {\bibinfo  {journal} {Surf. Coatings Technol.}\ }\textbf {\bibinfo
  {volume} {201}},\ \bibinfo {pages} {8388} (\bibinfo {year}
  {2007})}\BibitemShut {NoStop}%
\bibitem [{\citenamefont {Juslin}\ and\ \citenamefont
  {Nordlund}(2008)}]{Juslin2008}%
  \BibitemOpen
  \bibfield  {author} {\bibinfo {author} {\bibfnamefont {N.}~\bibnamefont
  {Juslin}}\ and\ \bibinfo {author} {\bibfnamefont {K.}~\bibnamefont
  {Nordlund}},\ }\href {\doibase 10.1016/j.jnucmat.2008.08.029} {\bibfield
  {journal} {\bibinfo  {journal} {J. Nucl. Mater.}\ }\textbf {\bibinfo {volume}
  {382}},\ \bibinfo {pages} {143} (\bibinfo {year} {2008})}\BibitemShut
  {NoStop}%
\bibitem [{\citenamefont {Karolewski}(2012)}]{Karolewski2012}%
  \BibitemOpen
  \bibfield  {author} {\bibinfo {author} {\bibfnamefont {M.~A.}\ \bibnamefont
  {Karolewski}},\ }\href {\doibase 10.1080/10420150.2012.700517} {\bibfield
  {journal} {\bibinfo  {journal} {Radiat. Eff. Defects Solids}\ }\textbf
  {\bibinfo {volume} {167}},\ \bibinfo {pages} {666} (\bibinfo {year}
  {2012})}\BibitemShut {NoStop}%
\bibitem [{\citenamefont {Zinoviev}\ and\ \citenamefont
  {Nordlund}(2017)}]{Zinoviev2017}%
  \BibitemOpen
  \bibfield  {author} {\bibinfo {author} {\bibfnamefont {A.~N.}\ \bibnamefont
  {Zinoviev}}\ and\ \bibinfo {author} {\bibfnamefont {K.}~\bibnamefont
  {Nordlund}},\ }\href {\doibase 10.1016/j.nimb.2017.03.047} {\bibfield
  {journal} {\bibinfo  {journal} {Nucl. Instrum. Methods Phys. Res., Sect. B}\ }\textbf {\bibinfo {volume} {406}},\
  \bibinfo {pages} {511} (\bibinfo {year} {2017})}\BibitemShut {NoStop}%
\bibitem [{\citenamefont {Lehtola}(2019{\natexlab{a}})}]{Lehtola2019c}%
  \BibitemOpen
  \bibfield  {author} {\bibinfo {author} {\bibfnamefont {S.}~\bibnamefont
  {Lehtola}},\ }\href {\doibase 10.1002/qua.25968} {\bibfield  {journal}
  {\bibinfo  {journal} {Int. J. Quantum Chem.}\ }\textbf {\bibinfo {volume}
  {119}},\ \bibinfo {pages} {e25968} (\bibinfo {year} {2019}{\natexlab{a}})},\
  \Eprint {http://arxiv.org/abs/1902.01431} {arXiv:1902.01431} \BibitemShut
  {NoStop}%
\bibitem [{\citenamefont {Lehtola}(2019{\natexlab{b}})}]{Lehtola2019b}%
  \BibitemOpen
  \bibfield  {author} {\bibinfo {author} {\bibfnamefont {S.}~\bibnamefont
  {Lehtola}},\ }\href {\doibase 10.1002/qua.25944} {\bibfield  {journal}
  {\bibinfo  {journal} {Int. J. Quantum Chem.}\ }\textbf {\bibinfo {volume}
  {119}},\ \bibinfo {pages} {e25944} (\bibinfo {year} {2019}{\natexlab{b}})},\
  \Eprint {http://arxiv.org/abs/1810.11653} {arXiv:1810.11653} \BibitemShut
  {NoStop}%
\bibitem [{\citenamefont {Lehtola}\ \emph {et~al.}(2019)\citenamefont
  {Lehtola}, \citenamefont {Dimitrova},\ and\ \citenamefont
  {Sundholm}}]{Lehtola2019d}%
  \BibitemOpen
  \bibfield  {author} {\bibinfo {author} {\bibfnamefont {S.}~\bibnamefont
  {Lehtola}}, \bibinfo {author} {\bibfnamefont {M.}~\bibnamefont {Dimitrova}},
  \ and\ \bibinfo {author} {\bibfnamefont {D.}~\bibnamefont {Sundholm}},\
  }\href {\doibase 10.1080/00268976.2019.1597989} {\bibfield  {journal}
  {\bibinfo  {journal} {Mol. Phys.} doi: 10.1080/00268976.2019.1597989} (\bibinfo {year}
  {2019})},\ \Eprint {http://arxiv.org/abs/1812.06274} {arXiv:1812.06274}
  \BibitemShut {NoStop}%
\bibitem [{\citenamefont {Herbert}(2015)}]{Herbert2015}%
  \BibitemOpen
  \bibfield  {author} {\bibinfo {author} {\bibfnamefont {J.~M.}\ \bibnamefont
  {Herbert}},\ }in\ \href {\doibase 10.1002/9781118889886.ch8} {\emph {\bibinfo
  {booktitle} {Rev. Comput. Chem.}}},\ Vol.~\bibinfo {volume} {28}\ (\bibinfo
  {publisher} {John Wiley \& Sons, Inc.},\ \bibinfo {year} {2015})\ pp.\
  \bibinfo {pages} {391--517}\BibitemShut {NoStop}%
\bibitem [{\citenamefont {Lehtola}(2019{\natexlab{c}})}]{Lehtola2019f}%
  \BibitemOpen
  \bibfield  {author} {\bibinfo {author} {\bibfnamefont {S.}~\bibnamefont
  {Lehtola}},\ }\href {\doibase 10.1063/1.5139948} {\bibfield  {journal}
  {\bibinfo  {journal} {J. Chem. Phys.}\ }\textbf {\bibinfo {volume} {151}},\
  \bibinfo {pages} {241102} (\bibinfo {year} {2019}{\natexlab{c}})},\ \Eprint
  {http://arxiv.org/abs/1911.10372} {arXiv:1911.10372} \BibitemShut {NoStop}%
\bibitem [{\citenamefont {Beebe}\ and\ \citenamefont
  {Linderberg}(1977)}]{Beebe1977}%
  \BibitemOpen
  \bibfield  {author} {\bibinfo {author} {\bibfnamefont {N.~H.~F.}\
  \bibnamefont {Beebe}}\ and\ \bibinfo {author} {\bibfnamefont
  {J.}~\bibnamefont {Linderberg}},\ }\href {\doibase
  http://dx.doi.org/10.1002/qua.560120408} {\bibfield  {journal} {\bibinfo
  {journal} {Int. J. Quant. Chem.}\ }\textbf {\bibinfo {volume} {12}},\
  \bibinfo {pages} {683} (\bibinfo {year} {1977})}\BibitemShut {NoStop}%
\bibitem [{\citenamefont {Folkestad}\ \emph {et~al.}(2019)\citenamefont
  {Folkestad}, \citenamefont {Kj{\o}nstad},\ and\ \citenamefont
  {Koch}}]{Folkestad2019}%
  \BibitemOpen
  \bibfield  {author} {\bibinfo {author} {\bibfnamefont {S.~D.}\ \bibnamefont
  {Folkestad}}, \bibinfo {author} {\bibfnamefont {E.~F.}\ \bibnamefont
  {Kj{\o}nstad}}, \ and\ \bibinfo {author} {\bibfnamefont {H.}~\bibnamefont
  {Koch}},\ }\href {\doibase 10.1063/1.5083802} {\bibfield  {journal} {\bibinfo
   {journal} {J. Chem. Phys.}\ }\textbf {\bibinfo {volume} {150}},\ \bibinfo
  {pages} {194112} (\bibinfo {year} {2019})},\ \Eprint
  {http://arxiv.org/abs/1811.12890} {arXiv:1811.12890} \BibitemShut {NoStop}%
\bibitem [{\citenamefont {Feng}\ \emph {et~al.}(2019)\citenamefont {Feng},
  \citenamefont {Epifanovsky}, \citenamefont {Gauss},\ and\ \citenamefont
  {Krylov}}]{Feng2019}%
  \BibitemOpen
  \bibfield  {author} {\bibinfo {author} {\bibfnamefont {X.}~\bibnamefont
  {Feng}}, \bibinfo {author} {\bibfnamefont {E.}~\bibnamefont {Epifanovsky}},
  \bibinfo {author} {\bibfnamefont {J.}~\bibnamefont {Gauss}}, \ and\ \bibinfo
  {author} {\bibfnamefont {A.~I.}\ \bibnamefont {Krylov}},\ }\href {\doibase
  10.1063/1.5100022} {\bibfield  {journal} {\bibinfo  {journal} {J. Chem.
  Phys.}\ }\textbf {\bibinfo {volume} {151}},\ \bibinfo {pages} {014110}
  (\bibinfo {year} {2019})}\BibitemShut {NoStop}%
\bibitem [{\citenamefont {Aquilante}\ \emph {et~al.}(2011)\citenamefont
  {Aquilante}, \citenamefont {Boman}, \citenamefont {Bostr{\"{o}}m},
  \citenamefont {Koch}, \citenamefont {Lindh}, \citenamefont {de~Mer{\'{a}}s},\
  and\ \citenamefont {Pedersen}}]{Aquilante2011}%
  \BibitemOpen
  \bibfield  {author} {\bibinfo {author} {\bibfnamefont {F.}~\bibnamefont
  {Aquilante}}, \bibinfo {author} {\bibfnamefont {L.}~\bibnamefont {Boman}},
  \bibinfo {author} {\bibfnamefont {J.}~\bibnamefont {Bostr{\"{o}}m}}, \bibinfo
  {author} {\bibfnamefont {H.}~\bibnamefont {Koch}}, \bibinfo {author}
  {\bibfnamefont {R.}~\bibnamefont {Lindh}}, \bibinfo {author} {\bibfnamefont
  {A.~S.}\ \bibnamefont {de~Mer{\'{a}}s}}, \ and\ \bibinfo {author}
  {\bibfnamefont {T.~B.}\ \bibnamefont {Pedersen}},\ }in\ \href {\doibase
  10.1007/978-90-481-2853-2_13} {\emph {\bibinfo {booktitle} {Linear-Scaling
  Tech. Comput. Chem. Phys.}}},\ Vol.~\bibinfo {volume} {13}\ (\bibinfo
  {publisher} {Springer Netherlands},\ \bibinfo {address} {Dordrecht},\
  \bibinfo {year} {2011})\ pp.\ \bibinfo {pages} {301--343}\BibitemShut
  {NoStop}%
\bibitem [{\citenamefont {Aquilante}\ \emph {et~al.}(2006)\citenamefont
  {Aquilante}, \citenamefont {Pedersen}, \citenamefont {{S{\'{a}}nchez de
  Mer{\'{a}}s}},\ and\ \citenamefont {Koch}}]{Aquilante2006}%
  \BibitemOpen
  \bibfield  {author} {\bibinfo {author} {\bibfnamefont {F.}~\bibnamefont
  {Aquilante}}, \bibinfo {author} {\bibfnamefont {T.~B.}\ \bibnamefont
  {Pedersen}}, \bibinfo {author} {\bibfnamefont {A.}~\bibnamefont
  {{S{\'{a}}nchez de Mer{\'{a}}s}}}, \ and\ \bibinfo {author} {\bibfnamefont
  {H.}~\bibnamefont {Koch}},\ }\href {\doibase 10.1063/1.2360264} {\bibfield
  {journal} {\bibinfo  {journal} {J. Chem. Phys.}\ }\textbf {\bibinfo {volume}
  {125}},\ \bibinfo {pages} {174101} (\bibinfo {year} {2006})}\BibitemShut
  {NoStop}%
\bibitem [{\citenamefont {Vahtras}\ \emph {et~al.}(1993)\citenamefont
  {Vahtras}, \citenamefont {Alml{\"{o}}f},\ and\ \citenamefont
  {Feyereisen}}]{Vahtras1993}%
  \BibitemOpen
  \bibfield  {author} {\bibinfo {author} {\bibfnamefont {O.}~\bibnamefont
  {Vahtras}}, \bibinfo {author} {\bibfnamefont {J.}~\bibnamefont
  {Alml{\"{o}}f}}, \ and\ \bibinfo {author} {\bibfnamefont {M.~W.}\
  \bibnamefont {Feyereisen}},\ }\href {\doibase 10.1016/0009-2614(93)89151-7}
  {\bibfield  {journal} {\bibinfo  {journal} {Chem. Phys. Lett.}\ }\textbf
  {\bibinfo {volume} {213}},\ \bibinfo {pages} {514} (\bibinfo {year}
  {1993})}\BibitemShut {NoStop}%
\bibitem [{\citenamefont {Aquilante}\ \emph {et~al.}(2007)\citenamefont
  {Aquilante}, \citenamefont {Lindh},\ and\ \citenamefont
  {Pedersen}}]{Aquilante2007a}%
  \BibitemOpen
  \bibfield  {author} {\bibinfo {author} {\bibfnamefont {F.}~\bibnamefont
  {Aquilante}}, \bibinfo {author} {\bibfnamefont {R.}~\bibnamefont {Lindh}}, \
  and\ \bibinfo {author} {\bibfnamefont {T.~B.}\ \bibnamefont {Pedersen}},\
  }\href {\doibase 10.1063/1.2777146} {\bibfield  {journal} {\bibinfo
  {journal} {J. Chem. Phys.}\ }\textbf {\bibinfo {volume} {127}},\ \bibinfo
  {pages} {114107} (\bibinfo {year} {2007})}\BibitemShut {NoStop}%
\bibitem [{\citenamefont {Aquilante}\ \emph {et~al.}(2009)\citenamefont
  {Aquilante}, \citenamefont {Gagliardi}, \citenamefont {Pedersen},\ and\
  \citenamefont {Lindh}}]{Aquilante2009}%
  \BibitemOpen
  \bibfield  {author} {\bibinfo {author} {\bibfnamefont {F.}~\bibnamefont
  {Aquilante}}, \bibinfo {author} {\bibfnamefont {L.}~\bibnamefont
  {Gagliardi}}, \bibinfo {author} {\bibfnamefont {T.~B.}\ \bibnamefont
  {Pedersen}}, \ and\ \bibinfo {author} {\bibfnamefont {R.}~\bibnamefont
  {Lindh}},\ }\href {\doibase 10.1063/1.3116784} {\bibfield  {journal}
  {\bibinfo  {journal} {J. Chem. Phys.}\ }\textbf {\bibinfo {volume} {130}},\
  \bibinfo {pages} {154107} (\bibinfo {year} {2009})}\BibitemShut {NoStop}%
\bibitem [{\citenamefont {Harbrecht}\ \emph {et~al.}(2012)\citenamefont
  {Harbrecht}, \citenamefont {Peters},\ and\ \citenamefont
  {Schneider}}]{Harbrecht2012}%
  \BibitemOpen
  \bibfield  {author} {\bibinfo {author} {\bibfnamefont {H.}~\bibnamefont
  {Harbrecht}}, \bibinfo {author} {\bibfnamefont {M.}~\bibnamefont {Peters}}, \
  and\ \bibinfo {author} {\bibfnamefont {R.}~\bibnamefont {Schneider}},\ }\href
  {\doibase 10.1016/j.apnum.2011.10.001} {\bibfield  {journal} {\bibinfo
  {journal} {Appl. Numer. Math.}\ }\textbf {\bibinfo {volume} {62}},\ \bibinfo
  {pages} {428} (\bibinfo {year} {2012})}\BibitemShut {NoStop}%
\bibitem [{\citenamefont {Millam}\ and\ \citenamefont
  {Scuseria}(1997)}]{Millam1997}%
  \BibitemOpen
  \bibfield  {author} {\bibinfo {author} {\bibfnamefont {J.~M.}\ \bibnamefont
  {Millam}}\ and\ \bibinfo {author} {\bibfnamefont {G.~E.}\ \bibnamefont
  {Scuseria}},\ }\href {\doibase 10.1063/1.473579} {\bibfield  {journal}
  {\bibinfo  {journal} {J. Chem. Phys.}\ }\textbf {\bibinfo {volume} {106}},\
  \bibinfo {pages} {5569} (\bibinfo {year} {1997})}\BibitemShut {NoStop}%
\bibitem [{\citenamefont {Challacombe}(1999)}]{Challacombe1999}%
  \BibitemOpen
  \bibfield  {author} {\bibinfo {author} {\bibfnamefont {M.}~\bibnamefont
  {Challacombe}},\ }\href {\doibase 10.1063/1.477969} {\bibfield  {journal}
  {\bibinfo  {journal} {J. Chem. Phys.}\ }\textbf {\bibinfo {volume} {110}},\
  \bibinfo {pages} {2332} (\bibinfo {year} {1999})}\BibitemShut {NoStop}%
\bibitem [{\citenamefont {Shao}\ \emph {et~al.}(2003)\citenamefont {Shao},
  \citenamefont {Saravanan}, \citenamefont {Head-Gordon},\ and\ \citenamefont
  {White}}]{Shao2003}%
  \BibitemOpen
  \bibfield  {author} {\bibinfo {author} {\bibfnamefont {Y.}~\bibnamefont
  {Shao}}, \bibinfo {author} {\bibfnamefont {C.}~\bibnamefont {Saravanan}},
  \bibinfo {author} {\bibfnamefont {M.}~\bibnamefont {Head-Gordon}}, \ and\
  \bibinfo {author} {\bibfnamefont {C.~A.}\ \bibnamefont {White}},\ }\href
  {\doibase 10.1063/1.1558476} {\bibfield  {journal} {\bibinfo  {journal} {J.
  Chem. Phys.}\ }\textbf {\bibinfo {volume} {118}},\ \bibinfo {pages} {6144}
  (\bibinfo {year} {2003})}\BibitemShut {NoStop}%
\bibitem [{\citenamefont {Lehtola}\ \emph {et~al.}(2012)\citenamefont
  {Lehtola}, \citenamefont {Hakala}, \citenamefont {Sakko},\ and\ \citenamefont
  {H{\"{a}}m{\"{a}}l{\"{a}}inen}}]{Lehtola2012}%
  \BibitemOpen
  \bibfield  {author} {\bibinfo {author} {\bibfnamefont {J.}~\bibnamefont
  {Lehtola}}, \bibinfo {author} {\bibfnamefont {M.}~\bibnamefont {Hakala}},
  \bibinfo {author} {\bibfnamefont {A.}~\bibnamefont {Sakko}}, \ and\ \bibinfo
  {author} {\bibfnamefont {K.}~\bibnamefont {H{\"{a}}m{\"{a}}l{\"{a}}inen}},\
  }\href {\doibase 10.1002/jcc.22987} {\bibfield  {journal} {\bibinfo
  {journal} {J. Comput. Chem.}\ }\textbf {\bibinfo {volume} {33}},\ \bibinfo
  {pages} {1572} (\bibinfo {year} {2012})}\BibitemShut {NoStop}%
\bibitem [{\citenamefont {Lehtola}(2018{\natexlab{a}})}]{erkale}%
  \BibitemOpen
  \bibfield  {author} {\bibinfo {author} {\bibfnamefont {S.}~\bibnamefont
  {Lehtola}},\ }\href {https://github.com/susilehtola/erkale} {\enquote
  {\bibinfo {title} {{ERKALE -- HF/DFT from Hel}},}\ } (\bibinfo {year}
  {2018}{\natexlab{a}})\BibitemShut {NoStop}%
\bibitem [{\citenamefont {L{\"{o}}wdin}(1956)}]{Lowdin1956}%
  \BibitemOpen
  \bibfield  {author} {\bibinfo {author} {\bibfnamefont {P.-O.}\ \bibnamefont
  {L{\"{o}}wdin}},\ }\href {\doibase 10.1080/00018735600101155} {\bibfield
  {journal} {\bibinfo  {journal} {Adv. Phys.}\ }\textbf {\bibinfo {volume}
  {5}},\ \bibinfo {pages} {1} (\bibinfo {year} {1956})}\BibitemShut {NoStop}%
\bibitem [{\citenamefont {Lehtola}(2019{\natexlab{d}})}]{Lehtola2019a}%
  \BibitemOpen
  \bibfield  {author} {\bibinfo {author} {\bibfnamefont {S.}~\bibnamefont
  {Lehtola}},\ }\href {\doibase 10.1002/qua.25945} {\bibfield  {journal}
  {\bibinfo  {journal} {Int. J. Quantum Chem.}\ }\textbf {\bibinfo {volume}
  {119}},\ \bibinfo {pages} {e25945} (\bibinfo {year} {2019}{\natexlab{d}})},\
  \Eprint {http://arxiv.org/abs/1810.11651} {arXiv:1810.11651} \BibitemShut
  {NoStop}%
\bibitem [{\citenamefont {Lehtola}(2018{\natexlab{b}})}]{HelFEM}%
  \BibitemOpen
  \bibfield  {author} {\bibinfo {author} {\bibfnamefont {S.}~\bibnamefont
  {Lehtola}},\ }\href {http://github.com/susilehtola/HelFEM} {\enquote
  {\bibinfo {title} {{HelFEM -- Finite element methods for electronic structure
  calculations on small systems}},}\ } (\bibinfo {year}
  {2018}{\natexlab{b}})\BibitemShut {NoStop}%
\bibitem [{\citenamefont {Lehtola}(2019{\natexlab{e}})}]{Lehtola2019}%
  \BibitemOpen
  \bibfield  {author} {\bibinfo {author} {\bibfnamefont {S.}~\bibnamefont
  {Lehtola}},\ }\href {\doibase 10.1021/acs.jctc.8b01089} {\bibfield  {journal}
  {\bibinfo  {journal} {J. Chem. Theory Comput.}\ }\textbf {\bibinfo {volume}
  {15}},\ \bibinfo {pages} {1593} (\bibinfo {year} {2019}{\natexlab{e}})},\
  \Eprint {http://arxiv.org/abs/1810.11659} {arXiv:1810.11659} \BibitemShut
  {NoStop}%
\bibitem [{\citenamefont {Lehtola}(2020{\natexlab{a}})}]{Lehtola2019e}%
  \BibitemOpen
  \bibfield  {author} {\bibinfo {author} {\bibfnamefont {S.}~\bibnamefont
  {Lehtola}},\ }\href {\doibase 10.1103/PhysRevA.101.012516} {\bibfield
  {journal} {\bibinfo  {journal} {Phys. Rev. A}\ }\textbf {\bibinfo {volume}
  {101}},\ \bibinfo {pages} {012516} (\bibinfo {year} {2020}{\natexlab{a}})},\
  \Eprint {http://arxiv.org/abs/1908.02528} {arXiv:1908.02528} \BibitemShut
  {NoStop}%
\bibitem [{\citenamefont {Jensen}(2001)}]{Jensen2001}%
  \BibitemOpen
  \bibfield  {author} {\bibinfo {author} {\bibfnamefont {F.}~\bibnamefont
  {Jensen}},\ }\href {\doibase 10.1063/1.1413524} {\bibfield  {journal}
  {\bibinfo  {journal} {J. Chem. Phys.}\ }\textbf {\bibinfo {volume} {115}},\
  \bibinfo {pages} {9113} (\bibinfo {year} {2001})}\BibitemShut {NoStop}%
\bibitem [{\citenamefont {de~Castro}\ and\ \citenamefont
  {Jorge}(1998)}]{DeCastro1998}%
  \BibitemOpen
  \bibfield  {author} {\bibinfo {author} {\bibfnamefont {E.~V.~R.}\
  \bibnamefont {de~Castro}}\ and\ \bibinfo {author} {\bibfnamefont {F.~E.}\
  \bibnamefont {Jorge}},\ }\href {\doibase 10.1063/1.475959} {\bibfield
  {journal} {\bibinfo  {journal} {J. Chem. Phys.}\ }\textbf {\bibinfo {volume}
  {108}},\ \bibinfo {pages} {5225} (\bibinfo {year} {1998})}\BibitemShut
  {NoStop}%
\bibitem [{\citenamefont {Lehtola}(2020{\natexlab{b}})}]{Lehtola2020b}%
  \BibitemOpen
  \bibfield  {author} {\bibinfo {author} {\bibfnamefont {S.}~\bibnamefont
  {Lehtola}},\ }\href {http://arxiv.org/abs/2001.04224} {\  (\bibinfo {year}
  {2020}{\natexlab{b}})},\ \Eprint {http://arxiv.org/abs/2001.04224}
  {arXiv:2001.04224} \BibitemShut {NoStop}%
\bibitem [{\citenamefont {Bloch}(1929)}]{Bloch1929}%
  \BibitemOpen
  \bibfield  {author} {\bibinfo {author} {\bibfnamefont {F.}~\bibnamefont
  {Bloch}},\ }\href {\doibase 10.1007/BF01340281} {\bibfield  {journal}
  {\bibinfo  {journal} {Zeitschrift f{\"{u}}r Phys.}\ }\textbf {\bibinfo
  {volume} {57}},\ \bibinfo {pages} {545} (\bibinfo {year} {1929})}\BibitemShut
  {NoStop}%
\bibitem [{\citenamefont {Dirac}(1930)}]{Dirac1930}%
  \BibitemOpen
  \bibfield  {author} {\bibinfo {author} {\bibfnamefont {P.~A.~M.}\
  \bibnamefont {Dirac}},\ }\href {\doibase 10.1017/S0305004100016108}
  {\bibfield  {journal} {\bibinfo  {journal} {Math. Proc. Cambridge Philos.
  Soc.}\ }\textbf {\bibinfo {volume} {26}},\ \bibinfo {pages} {376} (\bibinfo
  {year} {1930})}\BibitemShut {NoStop}%
\bibitem [{\citenamefont {Vosko}\ \emph {et~al.}(1980)\citenamefont {Vosko},
  \citenamefont {Wilk},\ and\ \citenamefont {Nusair}}]{Vosko1980}%
  \BibitemOpen
  \bibfield  {author} {\bibinfo {author} {\bibfnamefont {S.~H.}\ \bibnamefont
  {Vosko}}, \bibinfo {author} {\bibfnamefont {L.}~\bibnamefont {Wilk}}, \ and\
  \bibinfo {author} {\bibfnamefont {M.}~\bibnamefont {Nusair}},\ }\href
  {\doibase 10.1139/p80-159} {\bibfield  {journal} {\bibinfo  {journal} {Can.
  J. Phys.}\ }\textbf {\bibinfo {volume} {58}},\ \bibinfo {pages} {1200}
  (\bibinfo {year} {1980})}\BibitemShut {NoStop}%
\bibitem [{Note1()}]{Note1}%
  \BibitemOpen
  Note1,\ \href@noop {} {}\bibinfo {note} {Kai Nordlund, private communication,
  2020.}\BibitemShut {Stop}%
\bibitem [{\citenamefont {Becquart}\ \emph {et~al.}(2011)\citenamefont
  {Becquart}, \citenamefont {Barthe},\ and\ \citenamefont {{De
  Backer}}}]{Becquart2011}%
  \BibitemOpen
  \bibfield  {author} {\bibinfo {author} {\bibfnamefont {C.~S.}\ \bibnamefont
  {Becquart}}, \bibinfo {author} {\bibfnamefont {M.~F.}\ \bibnamefont
  {Barthe}}, \ and\ \bibinfo {author} {\bibfnamefont {A.}~\bibnamefont {{De
  Backer}}},\ }\href {\doibase 10.1088/0031-8949/2011/T145/014048} {\bibfield
  {journal} {\bibinfo  {journal} {Phys. Scr.}\ }\textbf {\bibinfo {volume}
  {T145}},\ \bibinfo {pages} {014048} (\bibinfo {year} {2011})}\BibitemShut
  {NoStop}%
\bibitem [{\citenamefont {Sand}\ \emph {et~al.}(2014)\citenamefont {Sand},
  \citenamefont {Nordlund},\ and\ \citenamefont {Dudarev}}]{Sand2014}%
  \BibitemOpen
  \bibfield  {author} {\bibinfo {author} {\bibfnamefont {A.~E.}\ \bibnamefont
  {Sand}}, \bibinfo {author} {\bibfnamefont {K.}~\bibnamefont {Nordlund}}, \
  and\ \bibinfo {author} {\bibfnamefont {S.~L.}\ \bibnamefont {Dudarev}},\
  }\href {\doibase 10.1016/j.jnucmat.2014.06.007} {\bibfield  {journal}
  {\bibinfo  {journal} {J. Nucl. Mater.}\ }\textbf {\bibinfo {volume} {455}},\
  \bibinfo {pages} {207} (\bibinfo {year} {2014})}\BibitemShut {NoStop}%
\bibitem [{\citenamefont {Pyykk{\"{o}}}(2012{\natexlab{a}})}]{Pyykko2012}%
  \BibitemOpen
  \bibfield  {author} {\bibinfo {author} {\bibfnamefont {P.}~\bibnamefont
  {Pyykk{\"{o}}}},\ }\href {\doibase 10.1146/annurev-physchem-032511-143755}
  {\bibfield  {journal} {\bibinfo  {journal} {Annu. Rev. Phys. Chem.}\ }\textbf
  {\bibinfo {volume} {63}},\ \bibinfo {pages} {45} (\bibinfo {year}
  {2012}{\natexlab{a}})}\BibitemShut {NoStop}%
\bibitem [{\citenamefont {Pyykk{\"{o}}}(2012{\natexlab{b}})}]{Pyykko2012a}%
  \BibitemOpen
  \bibfield  {author} {\bibinfo {author} {\bibfnamefont {P.}~\bibnamefont
  {Pyykk{\"{o}}}},\ }\href {\doibase 10.1021/cr200042e} {\bibfield  {journal}
  {\bibinfo  {journal} {Chem. Rev.}\ }\textbf {\bibinfo {volume} {112}},\
  \bibinfo {pages} {371} (\bibinfo {year} {2012}{\natexlab{b}})}\BibitemShut
  {NoStop}%
\end{thebibliography}%

\end{document}